\documentclass[aps,superscriptaddress]{revtex4}

\usepackage{amsmath, amsthm, amsfonts, amssymb}
\usepackage{arydshln}
\usepackage{array}
\usepackage{graphicx}
\usepackage{rotating}
\usepackage{tikz}
\usetikzlibrary{shapes.geometric, patterns}
\usetikzlibrary{decorations.pathreplacing, angles, quotes}
\usetikzlibrary{arrows, decorations.markings}
\usetikzlibrary{plotmarks}
\tikzset{
    buffer/.style={
        draw,
        regular polygon
    }
}

\theoremstyle{plain}

\theoremstyle{definition}

\def\av#1{\langle#1\rangle}
\def\P{\mathbb{P}}
\def\Plinks{\mathbb{P}^{\,\rm links}}
\def\Pnodes{\mathbb{P}^{\,\rm nodes}}

\def\Sn{S_n([2,3];p)}

\def\Mt#1{M_{\triangle,#1}}
\def\Mq#1{M_{\square,#1}}
\def\Nt#1{N_{\triangle,#1}}
\def\Nq#1{N_{\square,#1}}
\def\Lt#1{L_{\triangle,#1}}
\def\Lq#1{L_{\square,#1}}
\def\Ct#1{C_{\triangle,#1}}
\def\Cq#1{C_{\square,#1}}

\begin{document}

\title{Stochastic and mixed flower graphs}
\author{C. Tyler Diggans}

\affiliation{Clarkson Center for Complex Systems Science (C$^3$S$^2$), Clarkson University, Potsdam, NY 13699} 
\affiliation{Department of Physics, Clarkson University, Potsdam, NY 13699} 
\affiliation{Air Force Research Laboratory: Information Directorate, Rome, NY 13441}
\author{Erik M. Bollt}
\affiliation{Clarkson Center for Complex Systems Science (C$^3$S$^2$), Clarkson University, Potsdam, NY 13699} 

\affiliation{Department of Electrical and Computer Engineering, Clarkson University, Potsdam, NY 13699} 
\author{Daniel ben-Avraham} 
\affiliation{Clarkson Center for Complex Systems Science (C$^3$S$^2$), Clarkson University, Potsdam, NY 13699} 
\affiliation{Department of Physics, Clarkson University, Potsdam, NY 13699} 

\begin{abstract}
Stochasticity is introduced to a well studied class of recursively grown graphs: $(u,v)$-flower nets, which have power-law degree distributions as well as small-world properties (when $u=1$). The stochastic variant interpolates between different (deterministic) flower graphs thus adding flexibility to the model. The random multiplicative  growth process involved, however, leads to a spread ensemble of networks with finite variance for the number of links, nodes, and loops.  Nevertheless,  the degree exponent and loopiness exponent attain unique values in the thermodynamic limit of infinitely large graphs.  We also study a class of mixed flower networks, closely related to the stochastic flowers, but which are grown recursively in a {\it deterministic} way.  The deterministic growth of mixed flower-nets eliminates ensemble spreads, and their recursive growth allows for exact analysis of their (uniquely defined) mixed  properties. 
\end{abstract}
\maketitle
\centerline{\em{Revised: \today}}

\section{Introduction}
Generative graph models that reliably display properties found in real-world networks created something of a renaissance in the study of complex networks around the beginning of the millennium~\cite{Watts98,Barabasi99,Newman01,Dorogovtsev02,Albert02, Newman03b}. Most of these initial models were built on the assumptions that the network had to be stochastic, produce a compact graph, and include some form of preferential attachment~\cite{Barabasi99,Bianconi01,KRL,Zhang06}. However, a subset of models did away with  randomness, relying instead on deterministic recursive rules for their generation~\cite{Barabasi01,Comellas02,DGM,CFR04,Rozenfeld06, Zhang07}, and more recently this approach has been revisited in~\cite{Sun13,Barrier16}. One important early model, the so-called DGM network, or alternatively the Pseudofractal Scale-free Web (PSW), became a baseline example and has been studied with respect to many network properties over the years, including loopiness~\cite{Rozenfeld04}, diffusion~\cite{Bollt05}, percolation~\cite{Rozenfeld07}, spectral properties~\cite{Xie16}, and minimum dominating sets~\cite{Shan17}. A generalization of this important network: $(u,v)$-flower nets, was introduced in~\cite{Rozenfeld06} and enabled the construction of pseudo-fractal networks with varying network properties. Of particular interest in~\cite{Rozenfeld06} was a linearly scaling self-similarity leading to an associated transfinite dimension, which led to a relabeling of the special small-world cases as {\em transfractals}.   A similar generalization was put forward around the same time in~\cite{Zhang06}, which focused exclusively on variations of the PSW. Recursive constructions have the advantage of allowing for exact analysis of the graphs' structural and dynamical properties but are overly rigid, as they lack stochasticity. A weighted version of the PSW was introduced in~\cite{Zhang10}, which did allow for asymmetry in edge weights. It is reasonable to ask whether some measure of stochasticity can be introduced without losing all of the beneficial analysis tools that have been developed for strictly recursive structures.

In this paper we introduce stochasticity to a recursive construction called the $(u,v)$-flowers~\cite{Rozenfeld06}, by mixing two (or more) deterministic rules in a random fashion. Because of the random multiplicative nature of the growth process, the resulting distributions for the number of links and sites display a small but finite variance (relative to the average) even in the thermodynamic limit of infinitely large graphs. Nevertheless, some structural characteristics, such as the degree-exponent, attain a sharp, well-defined value, and can now be tuned at will by varying the randomness parameters. We also explore an alternative  {\em deterministic} method for mixing the recursive rules, to construct (unique) graphs that mimic stochastic flowers, yet eliminate the ensemble spread while still allowing for exact analysis of their properties.

The rest of this paper is organized as follows.  A brief review of  $(u,v)$-flowers is provided in Section~\ref{Recursive Flowers}. The stochastic growth process that mixes randomly between different recursive rules is explained in Section~\ref{Stochastic Flowers}. This section includes also an analysis of the distributions of the number of links ($M$) and nodes ($N$) and their moments, as well as  of the degree exponent and the loopiness exponent of the graphs.  (The loopiness exponent $\alpha$ measures how $h_*$---the most likely length of cycles in the graph---scales with its order, $h_*\sim N^{\alpha}$~\cite{Rozenfeld04}.)  Mixed flowers and their generation by deterministic mixing of
recursive rules is described in Section~\ref{Deterministic}, and their size, order, and statistics of cycles is analyzed exactly.   We conclude with a summary and discussion of the results in Section~\ref{Conclusions}.

\section{Recursive Flower Nets}
\label{Recursive Flowers}
The $(u,v)$-flower net~\cite{Rozenfeld06}, whose $n$th generation is denoted by $F_n(u,v)$, is constructed by a recursive process.  Flower-nets model many real-world properties (e.g., power-law degree distributions, small world when $u=1$, hierarchical community structure), and they importantly allow for exact analysis of their structure and of dynamical processes on them. The simplest case, of $(u,v)=(1,2)$, also known as the DGM network, was first described by Dorogotsev et al.,~\cite{DGM} as a \textit{pseudofractal}. The more general subset of small world cases $(u=1)$ with generic $v$ values were later dubbed \textit{transfractals} in~\cite{Rozenfeld06}, due to a transfinite dimensionality  defined in terms of a self-similarity that scales linearly in the distance between nodes.

\begin{figure}[ht!]
\centering
\includegraphics[width=6in]{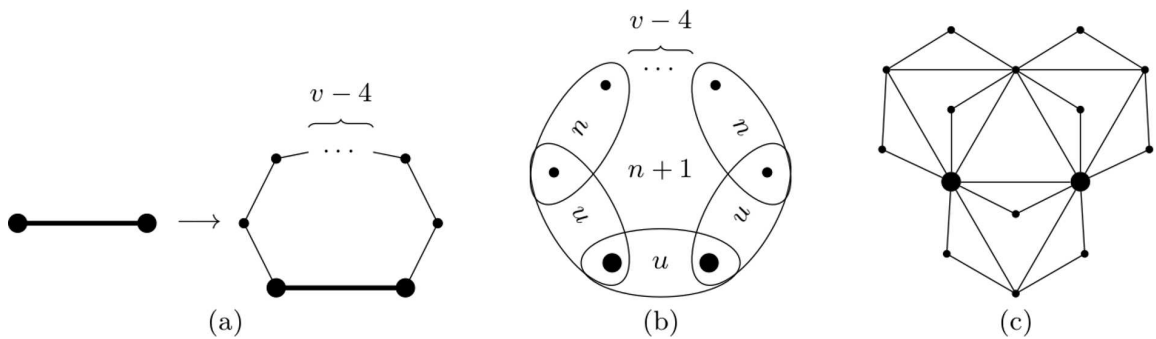}
\caption{The $(n+1)$-th generation of $F_n(1,v)$ is constructed by using either of two methods: (a)~Attach a path of $v$ links in parallel to each link of the previous generation, or (b)~paste $1+v$ copies of the $n$-th generation into a cycle, joining the copies hub to hub. (c)~$F_3(1,2)$, or the DGM network built to its third generation. The hubs are marked by larger full circles.}
\label{Pasting}
\end{figure}

The construction process begins (at generation $n=0$) with the complete graph on two nodes, $K_2$. To obtain the $(n+1)$-th generation, each link in generation $n$ is replaced by a parallel pair of paths of link lengths $u$ and $v$. This process is shown in Fig.~\ref{Pasting}(a) for the case of $u=1$. It was noted in~\cite{DGM} and later treatments~\cite{Rozenfeld06,Zhang07} that the $(n+1)$-th generation could also be obtained by pasting $w=u+v$ copies of generation $n$ at their hubs --- the two original nodes from generation $n=0$. This approach, illustrated in Fig.~\ref{Pasting}(b), is the basis to the  \textit{copy machine method} for generating  adjacency matrices (see Appendix C of~\cite{Bollt05}), and is useful for computational simulations. Fig.~\ref{Pasting}(c) provides an illustration of $F_3(1,2)$, or the DGM network built to its third generation.

The number of links, $M_n$, and nodes, $N_n$, in a flower graph of generation $n$ is~\cite{Bollt05,Rozenfeld07,Rozenfeld06}
\begin{gather}
\label{Mn}
	{M}_n = w^{n},\\
\label{Nn}
	{N}_n =\frac{w-2}{w-1} w^{n} + \frac{w}{w-1},
\end{gather}
where $w=u+v$ (thus, $w=3$ for the (1,2)-flower and $w=4$ for the (1,3)-flower).

A flower-graph of generation $n>0$ consists of nodes of degree $k=2,\dots,2^n$.  The number of nodes of degree $2^m$ in a flower graph of generation $n$ is
\begin{equation}	
\label{Nnm}
N_{2^m,n} = \left\{
	\begin{array}{ll}
		(w-2)w^{n-m}, & m<n,\\
		w,  & m=n.
	\end{array}\right. 
\end{equation}
In scale-free nets, in general, the number of nodes with degree larger or equal to $k$ scales as $N_{>k}\sim k^{1-\gamma}$.  According to~(\ref{Nnm}), $N_{>2^l}\sim w^{-l}$ (for $n\gg l\gg1$), so the degree exponent 
of flower graphs is
\begin{equation}
\label{gamma}
\gamma=1+\frac{\ln w}{\ln 2}\,.
\end{equation}

The loopiness exponent of $(1,v)$-flower graphs has been derived in~\cite{Rozenfeld04}:
\begin{equation}
\label{alpha}
\alpha=\frac{\ln v}{\ln w}\,,
\end{equation}
that is, $\alpha=\ln 2/\ln3$ for the $(1,2)$-flower, and $\alpha=\ln3/\ln4$ for the $(1,3)$-flower.
Finally, in $(1,2)$-flowers, the clustering coefficient~\cite{Watts98,Barmpoutis10} of nodes of degree $k=2^m$ is $C_k=2/k$.  Then, using the result from~(\ref{Nnm}), the average clustering coefficient for $F_n(1,2)$ is~\cite{DGM}
\begin{equation}
\label{clustering}
\av{C}_n=\frac{6^n+9}{2^{n-2}\cdot5(3^n+3)}\to\frac{4}{5}, {\ \rm as\ }n\to\infty\,.
\end{equation}
($(1,v)$-flowers with $v>2$ have zero clustering.)
Both the decay of $C_k$ with growing degree and the {\em finite} average of the overall clustering coefficient are characteristic properties of real-life complex networks.

\section{Stochastic Flower Graphs}
\label{Stochastic Flowers} 
Stochasticity is introduced to $(u,v)$-flower nets  so as to interpolate between the deterministic nets and form what we call \textit{stochastic flower graphs}. The increased flexibility gained in this way makes stochastic flower graphs better suited for modeling of real-life networks.  At the same time, they are largely amenable to the type of exact analysis that made flower nets so useful.  We focus on the {\em small-world} case (where $u=1$), to concur with most everyday life complex nets, but similar constructions could be envisioned for $u>1$ as well. 

\begin{figure}[ht!]
\centering
\includegraphics[width=6.5in]{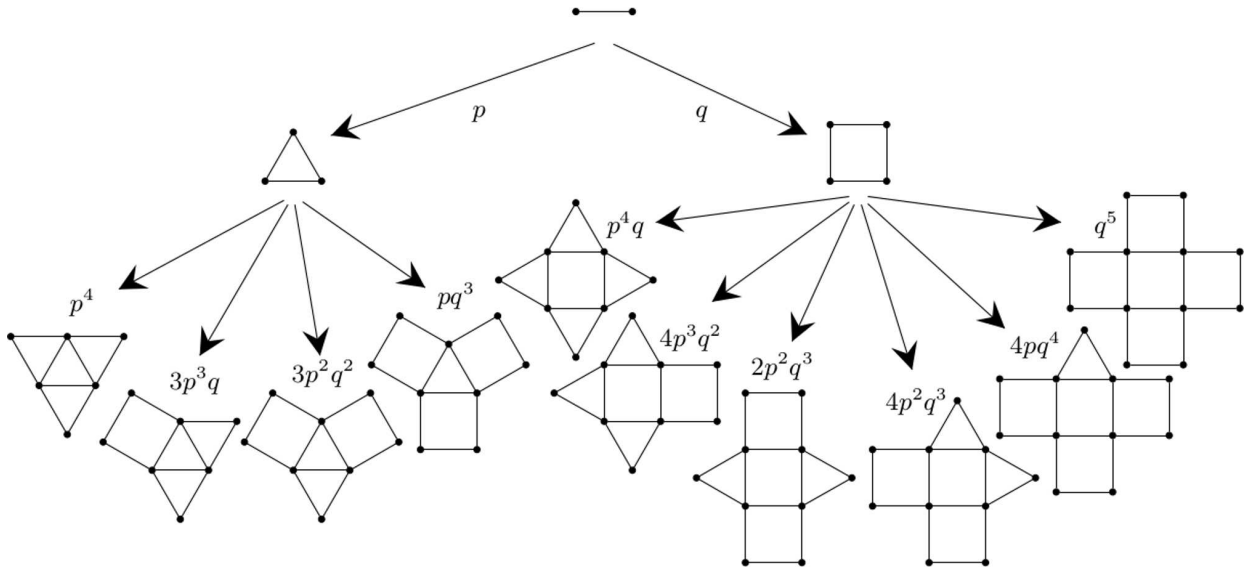}
\caption{Ensembles of $\Sn$, for $n=0,1,$ and $2$. The probability of obtaining any of the graphs in each ensemble is indicated. Notice the reduced symmetry for most members of the second generation.  The graphs become more disordered, or asymmetric, as $n$ gets larger.} 
\label{Stochastic}
\end{figure}

During the stochastic generation process, we vary $v$ on a link-by-link basis: each edge $e_i$ (in generation $n$) is replaced by two parallel paths of lengths 1 and $v_i$, or $(1,v_i)$-paths, where $v_i$ is selected from ${\rm\bf v}=[v_1,v_2,\dots]$ with probabilities ${\rm\bf p}=[p_1,p_2,\dots]$, respectively.  For simplicity, we restrict the present study to the simplest case of ${\rm\bf v}=[2,3]$ and ${\rm\bf p}=[p,1-p]$; in other words, each link is replaced by either $(1,2)$-paths (as in  the construction of the $(1,2)$-flower) with probability $p$, or by $(1,3)$-paths (as for the $(1,3)$-flower) with probability $1-p\equiv q$.   The first few generations of these $([2,3];p)$-stochastic flower graphs, denoted hereafter by $\Sn$, are shown in Fig.~\ref{Stochastic}.  One would expect the properties of $\Sn$ to interpolate smoothly between those of $(1,2)$-flowers an $(1,3)$-flowers, naively replacing $w$ in Eqs.~(\ref{Mn}) -- (\ref{gamma}) with the ``average"
$\bar w=3p+4q=3+q$.  We shall see that this naive expectation works quite well, to some extent.  For the loopiness exponent, however, naively replacing $v$ with $\bar v=2p+3q=2+q$ and $w$ with $\bar w$ in~(\ref{alpha}), fails to produce the right answer.

The hub-pasting approach of Fig.~\ref{Pasting}(b) works also for the stochastic flower graphs: to obtain generation $n+1$, simply paste either 3 graphs from generation $n$ (with probability $p$), or 4 graphs (with probability $q$).  The graphs to be pasted together are drawn randomly from the ensemble of graphs of generation $n$, with their respective probabilities.  It can be shown by induction that the link-by-link replacement and the pasting of subgraphs result in identical stochastic ensembles. The equivalence of these two methods of construction is important because it is often advantageous to use one form or the other for the analysis of  structural properties and dynamical processes by exact analytic recursions.

\subsection{Distributions of Links and Nodes}
\label{Moments}
Suppose $\Sn$ has $M$ links, with probability $\Plinks_n(M)$.  An iteration of the graph would then yield $3M+m$ links $(m=0,1,\dots,M)$, with probability $\binom{M}{m}p^{M-m}q^m$.  Thus,
\begin{equation}
\label{Mdistrib}
\Plinks_{n+1}(M) = \sum_{M'}\sum_{\fontsize{5pt}{5pt}\selectfont \begin{array}{cc}
m\\
m+3M'=M\\
\end{array}}{\Plinks_n(M') \binom{M'}{m} p^{M'-m} q^m}\,,
\end{equation}
where the outer sum runs over all possible values of $M'$ and the inner sum runs over all possible values of $m$ such that $3M'+m=M$  ($\Plinks_n=0$ for unattainable values of $M$).  One can then use Eq.~(\ref{Mdistrib}), along with $\Plinks_0(M)=\delta_{M,1}$, to generate all the $\Plinks_n(M)$ recursively.

Let $\av{M^r}_n=\sum_MM^r\Plinks_n(M)$ denote the $r$-th moment for the number of links in generation $n$.
It then follows from~(\ref{Mdistrib}) that 
\begin{equation}
\av{M^r}_{n+1} = \sum_{M}\sum_{m=0}^M{\Plinks_n(M) \binom{M}{m} p^{M-m} q^m (3M+m)^r}.
\end{equation}
Beginning with the first moment (setting $r=1$), we have
\begin{equation}
\begin{split}
\av{M}_{n+1}&=\sum_M\sum_{m=0}^M\Plinks_n(M)\binom{M}{m}p^{M-m}q^m(3M+m)\\
&=3\av{M}_n+\sum_M\Plinks_n(M)q\frac{\partial}{\partial q}(p+q)^M\\
&=(3+q)\av{M}_n\,,
\end{split}
\end{equation}
which along with the initial condition $\av{M}_0=1$, yields
\begin{equation}
\label{M}
\av{M}_n=(3+q)^n\,.
\end{equation}
This is identical to~(\ref{Mn}), obtained by the na\"{i}ve substitution.

For the second moment, similar manipulations lead to the recursion relation
\begin{equation}
\av{M^2}_{n+1} = (3+q)^2 \av{M^2}_{n} + pq (3+q)^n,
\end{equation}
and, along with $\av{M^2}_0 = 1$, we get
\begin{equation}
\label{M2}
\av{M^2}_{n} = \left(1 + \frac{pq}{(2+q)(3+q)}\right) (3+q)^{2n} - \frac{pq}{(2+q)(3+q)} (3+q)^n.
\end{equation}

The derivation of recursion relations for higher moments is generally cumbersome,
however, the first two moments suffice to explore the convergence of the distribution in the thermodynamic limit (of $n\to\infty$).  Indeed, the variance of the distribution for generation $n$, is
\[
\sigma_{M,n}^2=\av{M^2}_n-\av{M}^2_n=qp\frac{(3+q)^n((3+q)^n-1)}{(2+q)(3+q)}\,,
\]
so that
\begin{equation}
\label{mu}
\frac{\sigma_{M,n}^2}{\av{M}_n^2}=qp\frac{1-(3+q)^{-n}}{(2+q)(3+q)}\to \frac{qp}{(2+q)(3+q)}\equiv \mu(p)^2, \>{\rm as\ }n\to\infty\,.
\end{equation}
In other words, the distribution of the number of links, scaled to their average, $\Plinks_n(M/\av{M}_n)$, converges exponentially in $n$ (or equivalently, as $1/M_n)$. Notice that the standard deviation of the converged distribution is a \textit{non-vanishing} fraction of the average: for example, $p=0.5$ leads to $\mu(p)^2 = 1/35$. On the other hand, since the support of $\P_n(M)$ is $M\in [3^n, 4^n]$ and $\sigma_{M,n}\to \mu(p)(3+q)^n$, the standard deviation is a vanishing fraction of the support in the thermodynamic limit, and in that sense the distribution can be considered ``sharp." 

\begin{figure}[ht!]
\includegraphics[width=4.5in]{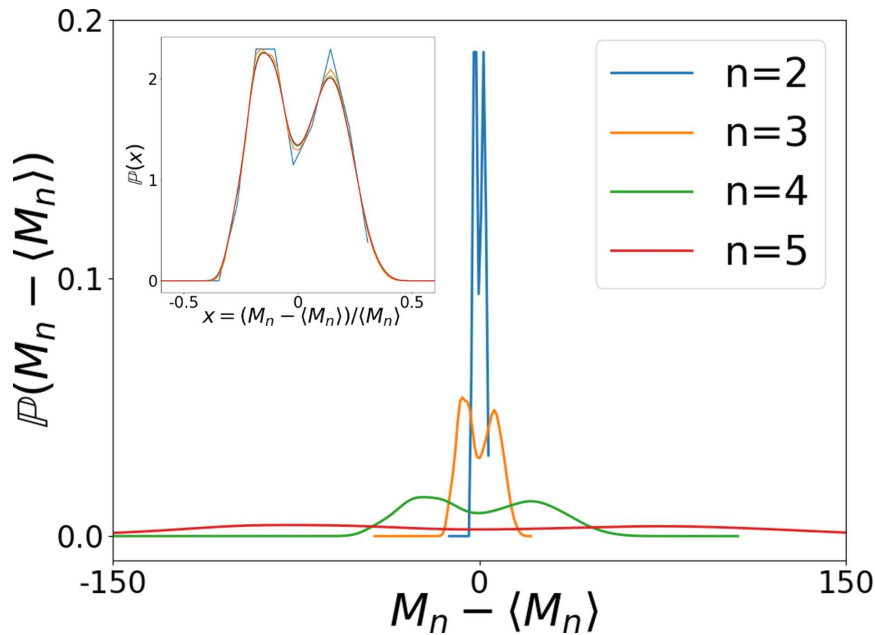}
\caption{(Color online) The distribution of links  $\Plinks(M)$ when $p=0.5$ for generations $n = 2$--$5$ plotted against  $M-\av{M}$. Inset: The scaled distribution $\Plinks((M-\av{M})/\av{M})$ quickly converges to a limit curve with $n$.}
\label{Sharp}
\end{figure}

Figure~\ref{Sharp} shows the distributions for the number of links  $\Plinks_n(M)$ in generations $2$ through $5$ for $p=0.5$.  The bimodal feature is an artifact of the two possibilities for $n=1$: the $(1,2)$-flower (with probability $p$) or the $(1,3)$-flower (probability $q$).  If one takes either of these two cases as the starting configuration (instead of $K_2$), the peaks reappear, due to the split in the next generation, but are now much closer together.  This trend continues as the initial seed network gets larger and larger. 

Working out the distribution for the number of nodes $\Pnodes_n(N)$ proves more difficult. In particular, beginning with a graph of $M_n$ links and $N_n$ nodes; if $m$ of the links are selected to generate a $(1,3)$-path (and the remaining $M_n-m$ to $(1,2)$-paths), then the resultant graph in generation $n+1$  consists of
\begin{equation}
\left\{
\begin{aligned}
M_{n+1} &= 3M_n + m\,,\\
N_{n+1} &= N_n + M_n + m\,,
\end{aligned}\right.
\end{equation}
links and nodes respectively. Thus, while links can be analyzed in closed form, as done above, the number of nodes, $N_n$, cannot be studied independently from the number of links, $M_n$.

A simple way to overcome this difficulty is to define a generating function for the joint probability distribution for the number of links and nodes in generation $n$, $\P_n(M,N)$, as: 
\begin{equation}
\label{joint}
f_n(x,y) = \sum_{M,N}\P(M,N) x^M y^N.
\end{equation}
For example, for the usual initial condition of $S_0([2,3];p)=K_2$, $\P_0(M,N)=\delta_{M,1}\delta_{N,2}$ and thus $f_0(x,y)=xy^2$.  Since each link that evolves into a $(1,2)$-path (with probability $p$) introduces two new links and one new node, while each link that evolves into a $(1,3)$-path (with probability $q=1-p$) introduces three new links and two new nodes, the generating function for the evolved graph can be obtained from the mapping:
$x\mapsto px\cdot x^2y+qx\cdot x^3y^2$, $y\mapsto y$ (the nodes, represented by $y$, do not evolve), and thus
\begin{equation}
\label{mapping}
f_{n+1}(x,y) = f_n(px^3y + qx^4y^2, y).
\end{equation}
The $f_n$'s can then be obtained by iterating this relation. 

From the joint probability distribution for links and nodes one can obtain the distribution for links or nodes alone:
\begin{equation}
\Plinks_n(M) = \sum_{N}{\P(M,N)}\quad ; \qquad\Pnodes_n(N) = \sum_{M}{\P(M,N)}.  
\end{equation}
The summation over $N$ can be effected by setting $y=1$, i.e.,  $f_{n+1}(x,1)=f_{n+1}^{\,{\rm links}}(x)=f_{n}^{\,{\rm links}}(px^3+qx^4)$, leading to the results for $\Plinks$ discussed above. Summing over $M$ in this way is more complex,  because due to the coupling between $M$ and $N$ the substitution of $x=1$ should be done only at the end, after having iterated the full expression with $x$ and $y$ to the $n$-th generation.  Instead, we derive equations for the moments of $N$.

Using Eqs.~(\ref{joint}) and~(\ref{mapping}), we have
\begin{equation}
\av{N}_{n+1} = y \frac{\partial}{\partial y} \sum_{M,N}{\P_n(M,N)(px^3y+qx^4y^2)^M y^N \rvert_{\genfrac{}{}{0pt}{2}{x=1}{y=1}}} = (1+q)\av{M}_n + \av{N}_n.
\end{equation}
Putting in the result from~(\ref{M}) for $\av{M}_n$, along with $\av{N}_0 = 2$, and solving for $\av{N}_n$, we find 
\begin{equation}
\label{Nn_s}
\av{N}_n = \frac{1+q}{2+q}(3+q)^n + \frac{3+q}{2+q},
\end{equation}
which again coincides with the result for $(u,v)$-flowers, Eq.~(\ref{Nn}), with the expected substitution $\bar w=(3+q)$. 

For the second moment, we begin similarly with 
\begin{equation}
\label{N2}
\begin{aligned}
\av{N^2}_{n+1} &= y \frac{\partial}{\partial y}y \frac{\partial}{\partial y} \sum_{M,N}{\P_n(M,N)(px^3y+qx^4y^2)^M y^N \rvert_{\genfrac{}{}{0pt}{2}{x=1}{y=1}}}\\
& =qp\av{M}_n +(1+q)^2\av{M^2}_n + 2 (1+q)\av{MN}_n + \av{N^2}_n.
\end{aligned}
\end{equation}
Thus, in order to solve for $\av{N^2}_n$, we require a formula for $\av{MN}_n$, which we get from an additional recursion relation:
\begin{equation}
\begin{aligned}
\av{MN}_{n+1} &= x \frac{\partial}{\partial x}y \frac{\partial}{\partial y} \sum_{M,N}{\P_n(M,N)(px^3y+qx^4y^2)^M y^N \rvert_{\genfrac{}{}{0pt}{2}{x=1}{y=1}}}\\
& =(3+q)\av{MN}_n +(1+q)(3+q)\av{M^2}_n + pq\av{M}_n.
\end{aligned}
\end{equation}
Starting with the initial condition, $\av{MN}_0 = 2$, and the results of Eqs.~(\ref{M}) and~(\ref{M2}) one can then solve explicitly for $\av{MN}_n$; armed with this result and the initial condition of $\av{N^2}_0 = 4$, one can then use~(\ref{N2}) to obtain an explicit solution for $\av{N^2}_n$. The actual expressions are lengthy and we instead present only the large$-n$ asymptotic limits:
\begin{equation}
\begin{aligned}
\av{MN}_n &\sim \frac{6(1+q)^2}{(2+q)^2(3+q)}(3+q)^{2n}, \text{ as } n\rightarrow\infty,\\
 \av{N^2}_n &\sim  \frac{6(1+q)^3}{(2+q)^3(3+q)}(3+q)^{2n}, \text{ as } n\rightarrow\infty.
\end{aligned}
\end{equation}
We conclude that, as in the case for links, $\Pnodes_n(N)$ too has a standard deviation proportional to its average:
\begin{equation}
\label{N_fluc}
\frac{\sigma^2_{N,n}}{\av{N}^2_n}\rightarrow \frac{qp}{(2+q)(3+q)}\equiv\nu(p)^2\,,
\end{equation}
and with the same fractional ratio, i.e., $\nu(p)=\mu(p)$  [Eq.~(\ref{mu})].

In what follows we show that despite the spread in the distributions of structural properties such as the number of links and nodes, characteristic exponents of stochastic flowers, such as the degree exponent and the loopiness exponent, attain perfectly well-defined sharp limits.

\subsection{Degree Exponent}
\label{Degree}
For a stochastic flower of generation $n$, $\Sn$, the degree sequence is still $\{k\}=\{1,2,2^2,\dots,2^n\}$, same as for $n$-generation $(u,v)$-flowers in general.  The average number of nodes with degree larger than $2^m$ scales like $N_{>2^m}\sim\bar w^{n-m}$.  However, as implied by the foregoing discussion on the distribution of $N$, the typical fluctuation in $N_{>2^m}$ is of the same order as its average, that is,
\begin{equation}
N_{>2^m}\sim (1+\sigma)\bar w^{n-m}\,,
\end{equation}
where $-\nu(p)\lesssim\sigma\lesssim \nu(p)$ is a random variable  [c.f. Eq.~(\ref{N_fluc})].  Thus, a plot of $\ln N_{>k}$ vs.~$\ln k$ would result in a curve that meanders typically between the two parallel lines of $-m\ln\bar w
+\ln(1\pm\nu(p))$ plotted against $m\ln 2$.   A linear fit to that curve, as $n\to\infty$, would therefore converge to the degree exponent
\begin{equation}
\label{gamma_s}
\gamma=1+\frac{\log\bar w}{\log2}=1+\frac{\ln(3+q)}{\ln2}\,,
\end{equation}
with an error that vanishes as $1/n$ (or more precisely, as $\ln[(1+\nu)/(1-\nu)]/n\ln2$).  Thus, the degree exponent  $\gamma$ achieves a well-defined value in the thermodynamic limit, despite the non-vanishing fluctuations in $N$.

\subsection{Loopiness and Clustering}
\label{Loopiness}
The \textit{loopiness} exponent $\alpha$ characterizes how the most likely length  of a cycle in the graph, $h_*$, scales with its order (number of nodes, $N$)~\cite{Rozenfeld04}: 
\begin{equation}
h_* \sim N^\alpha.
\end{equation}
The statistics of loops was also explored for two types of random scale-free graphs in~\cite{Bianconi05}, and subsequent work explored efficient algorithms for obtaining the statistics of cycles in generic graphs~\cite{Klemm06},~\cite{Marinari07}. 
Following the development in~\cite{Rozenfeld04}, we denote the number of loops, or cycles of length $h$ in a graph of generation $n$, by $C_n(h)$; and the number of paths of length $h$ connecting the two hub nodes in generation $n$, by $L_n(h)$.  Using the recursive construction of pasting copies of generation $n$ to produce generation $n+1$, one can write exact recursion relations for these quantities. In a $(1,2)$-flower, for example,
\begin{equation}
\begin{split}
\label{cyclesFn}
L_{n+1}(h)&=L_n(h) + \sum_{j+k=h}L_n(j)L_n(k)\,,\\
C_{n+1}(h)&=3C_n(h) + \sum_{j+k+l=h}L_n(j)L_n(k)L_n(l)\,.
\end{split}
\end{equation}
The first equation expresses the fact that a path between the two hubs in generation $n+1$ could go through the single subunit of generation $n$, or span the other two $``v"$-subunits; similarly, cycles of length $h$ could be found in each of the three subunits, or made up of three paths spanning the generation-$n$ subunits across their hubs (second equation). 
Once again, the generating functions
 \begin{equation}
 L_n(z) = \sum_h {L_n(h)} z^h\,,\qquad C_n(z) = \sum_h C_n(h) z^h\,, 
 \end{equation}
 simplify the analysis, as the convolution terms become then simple products.  Applying these ideas to the $\Sn$ stochastic flowers, we get
\begin{equation}
\label{cyclesSn}
\left\{\begin{aligned}
L_{n+1}(z) & = L_{n}(z) + p L_{n}^2(z) + q L_{n}^3(z)\\
C_{n+1}(z) &= (3+q) C_n(z) + p L_{n}^3(z) + q L_{n}^4(z)\\
\end{aligned}\right. .
\end{equation}
The first line, in this case, denotes the fact that apart from spanning a path through the single $n$-subunit, the $(n+1)$-path could go through either two $v$-subunits (with probability $p$), or three $v$-subunits (prob. $q=1-p$), depending on which $v$ was selected in the recursive construction; and similarly for the second line. 

\begin{figure}[ht!]
\includegraphics[width=4.5in]{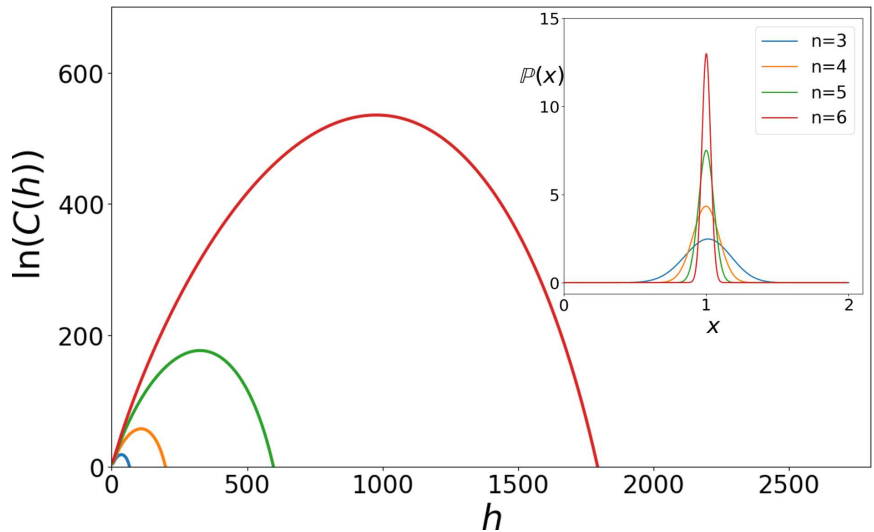}
\caption{(Color online) Plot of $\ln C_n(h)$ vs. $h$ for generations $n=3,4,5$ and $6$ in stochastic flower graphs with $p=1/2$. The inset shows $\P(h/h_*)$ --- the probability distribution density of $h$-cycles scaled to their most likely length, $h_*$.}
\label{AveLoopiness}
\end{figure}

Starting with the initial condition $L_0(z)=z$ and $C_0(z)=0$, Eqs.~(\ref{cyclesSn}) can be iterated and $C_n(h)$ can then be obtained from the coefficient of $z^h$ in $C_n(z)$.  In Fig.~\ref{AveLoopiness} we plot the statistics of cycles obtained in this way for generations $n=3,4,5,6$ and $p=1/2$.  The results show that the most probable cycle length $h_*$ grows by a factor of 3 from one generation to the next. That is also true for other values of $p<1$ (not shown).  Since $h_*\sim3^n$ and $N_n\sim\bar w^n$, we conclude that the loopiness exponent of $\Sn$ is
\begin{equation}
\label{alphaSn}
\alpha=\frac{\ln{3}}{\ln{\bar w}}=\frac{\ln{3}}{\ln{(3+q)}}\,,\quad p<1\,.
\end{equation}
(For the deterministic case of $p=1$, one obtains $(1,2)$-flowers, with $\alpha=\ln 2/\ln 3$.)  We see that naively replacing $v$ with $\bar v=2p+3q$ in~(\ref{alpha}) does not work, and instead any {\em finite} $q$
yields a loopiness exponent characteristic of $(1,3)$-flowers.

Note that while the recursions~(\ref{cyclesFn}) for $F_n(1,2)$ are exact, the same is {\em not} true for the analogous recursions~(\ref{cyclesSn}) for the $\Sn$.  The problem is that $L_n(h)$ and $C_n(h)$ in this case do not have a single sharp value (as for deterministic flowers), but represent a distribution of values.  Furthermore, the replacement of the distributions by their average does not work, as these distributions do not converge to delta-functions as $n\to\infty$,  and the equations are nonlinear (in general, $\av{f(L_n)}\neq f(\av{L_n})$).  Nevertheless, the naive approach followed here, of replacing the distributions by their average, yields the correct result~(\ref{alphaSn}) for $\alpha$, as shown in the next section.  

Clustering in stochastic flowers can be dealt with, at least qualitatively, by following the same procedure of replacing distributions with their averages.  Following this approach for $\Sn$, we find
\begin{equation}
\av{C}_n=\frac{p}{2^{n-1}}\frac{\bar w-1}{(\bar w-2)\bar w^n+\bar w}\left[\bar w+\frac{\bar w-2}{2\bar w-1}\left((2\bar w)^n-2\bar w\right)\right],
\end{equation}
and in the limit of $n\to\infty$
\begin{equation}
\av{C}_\infty=\frac{2p(\bar w-1)}{2\bar w-1}=\frac{2p(3-p)}{7-2p}\,.
\end{equation}
Thus, $\Sn$ retains the feature of finite average clustering in the thermodynamic limit.  As expected, $\av{C}_\infty$ interpolates between 0 for $p=0$, or $F(1,3)$, to $4/5$ for $p=1$, or $F(1,2)$.

To better address the tricky issues arising from ensemble spread, we next introduce a class of {\em mixed} flower-graphs that are fully deterministic, yet closely mimic typical members of the stochastic ensemble of $\Sn$.

\section{Mixed Flower Graphs}
\label{Deterministic}
\begin{figure}[ht!]
\centering
\includegraphics[width=6in]{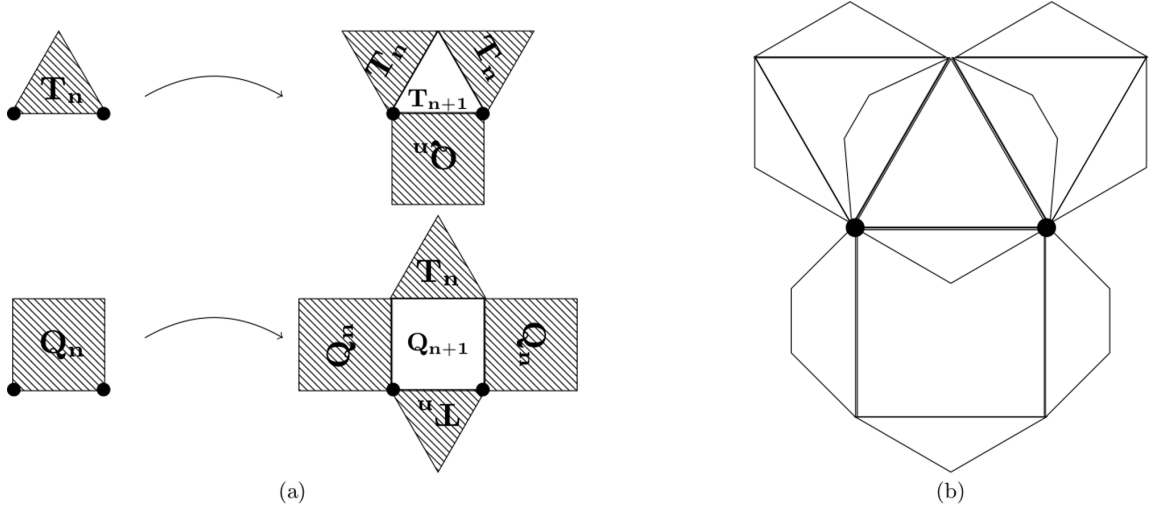}
\caption{(a)~Generation of $T_{n+1}$ and $Q_{n+1}$ from $T_n$'s and $Q_n$'s by the rule set  $\{QTT;TQTQ\}$.  The hubs are indicated by large solid circles.  (b)~$T_3$, obtained by the rule set $\{QTT;TQTQ\}$.}
\label{Mixed}
\end{figure}

We now introduce a class of {\em mixed flower nets} whose recursive construction is completely deterministic; there is a {\em single} instance of each mixed flower net.  The recursive construction generates two networks in tandem: $T_n$, that starts from an initial 3-cycle seed, $T_1$, in generation $n=1$, and $Q_n$, that starts from a 4-cycle seed, $Q_1$.  $T_{n+1}$ and $Q_{n+1}$ are obtained by pasting 3 or 4 networks of generation $n$ at their hubs, respectively, according to some prescribed rule.  Figure~\ref{Mixed}(a) shows a specific example of such a rule: $T_{n+1}$ is obtained by pasting $Q_n,T_n,T_n$ together at the hubs, starting with $Q_n$ and proceeding counterclockwise.  The hubs of $T_{n+1}$ are the hubs of the first subnet in the sequence ($Q_n$, in this case).  Similarly, $Q_{n+1}$ is obtained by pasting $T_n,Q_n,T_n,Q_n$ at the hubs, starting with $T_n$ and proceeding counterclockwise. The hubs of $Q_{n+1}$ are the hubs of the first subnet in the sequence ($T_n$).  This particular set of rules can then be denoted more succinctly as $\{QTT;TQTQ\}$. Figure~\ref{Mixed}(b) shows $T_3$, obtained by this rule set.  Since mixed flowers are deterministic constructs, their structural properties can be obtained from {\em exact} recursion relations, as shown by the few examples below.

\subsection{Size and Order of Mixed Flower Graphs}
Focusing, as an example, on the mixed flower nets resulting from the  rule set $\{QTT;TQTQ\}$, the number of links in $T_n$ ($\Mt{n}$) and in $Q_n$ ($\Mq{n}$) obey the recursion relation
\begin{equation}
\label{MtMq}
\begin{aligned}
\left[\begin{aligned}
\Mt{n+1} \\
\Mq{n+1}
\end{aligned}\right] 
&= \left[\begin{array}{cc}
2 & 1 \\ 
2 & 2 \end{array}\right] \left[\begin{aligned}
\Mt{n} \\
\Mq{n}
\end{aligned}\right].
\end{aligned}
\end{equation}
This, together with the initial conditions $\Mt{1}=3$, $\Mq{1}=4$, yields
\begin{equation}
\begin{aligned}
M_{\triangle,n} &= \left(\frac{3}{2}+\sqrt{2}\right) r_+^{n-1} + \left(\frac{3}{2}-\sqrt{2}\right) r_-^{n-1}\,,\\ 
M_{\square,n} &= \left(2+\frac{3\sqrt{2}}{2}\right) r_+^{n-1} + \left(2-\frac{3\sqrt{2}}{2}\right) r_-^{n-1}\,,
\end{aligned}
\end{equation}
where 
\[
r_{\pm}=2\pm\sqrt{2}
\]
are the eigenvalues of the  matrix ${\rm\bf A}={{2\, 1}\choose{2 \,2}}$.  For the number of nodes, $\Nt{n}$ and $\Nq{n}$, we have

\begin{equation}
\label{NtNq}
\begin{aligned}
\left[\begin{aligned}
\Nt{n+1} \\
\Nq{n+1}
\end{aligned}\right] 
&= \left[\begin{array}{cc}
2 & 1 \\ 
2 & 2 \end{array}\right] \left[\begin{aligned}
\Nt{n} \\
\Nq{n}
\end{aligned}\right]
-\left[\begin{aligned}
3 \\
4
\end{aligned}\right],
\end{aligned}
\end{equation}
the difference being that here we subtract the nodes identified in the pasting of the subgraphs, to avoid over-counting.  In view of the initial conditions $\Nt{1}=3$, $\Nq{1}=4$, the solutions are
\begin{equation}
\begin{aligned}
N_{\triangle,n} &=\frac{1}{2}(r_+^n + r_-^n)+1\,,\\
N_{\square,n} &= (1+\sqrt{2})r_+^{n-1} + (1-\sqrt{2})r_-^{n-1}+2\,.
\end{aligned}
\end{equation}

We see that the growth of links and nodes in all cases is dominated by the larger eigenvalue of {\bf A}, $r_+$:
$\Mt{n}\sim\Mq{n}\sim\Nt{n}\sim\Nq{n}\sim r_+^n$, as $n\to\infty$.  Comparing this observation with the result for stochastic flowers, $\av{M}_n\sim\av{N}_n\sim\bar w^n$; $\bar w=3+q$, we deduce that the rules of growth in our example result in a (deterministic) mixed flower (either $T_n$ or $Q_n$) that mimics stochastic flowers with $p=2-\sqrt{2}\equiv\tilde p\approx0.5858$, $S_n([2,3];2-\sqrt{2})$.  From a different perspective, $\tilde p\Mt{n}+\tilde q\Mq{n}\equiv\bar M_n$ and $\tilde p\Nt{n}+\tilde q\Nq{n}\equiv \bar N_n$ are $L_1$-norms of eigenvectors of {\bf A} corresponding to $r_+$, and
$\bar M_n$ and $\bar N_n$ are identical to $\av{M}_n$ and $\av{N}_n$ of Eqs.~(\ref{M}) and (\ref{Nn_s}), for $p=\tilde p$.

\subsection{Degree Exponent of Mixed Flowers}

The degree sequence of both $T_n$ and $Q_n$ is $\{k\}=\{2,2^2,\dots,2^n\}$, same as for $(u,v)$-flower nets and stochastic flowers in general.  The number of nodes of degree $k=2^m$ scales as
\[
N_{2^m,n}=N_{n-m+1}-N_{n-m}\sim r_+^{n-m}\,,
\]
according to~(\ref{NtNq}).  Therefore, $N_{>2^m}\sim r_+^{-m}\sim (2^m)^{1-\gamma}$, for $1\ll m\ll n$, and 
\begin{equation}
\gamma=1+\frac{\ln\tilde w}{\ln 2}\,,
\end{equation}
where $\tilde w=r_+=3\tilde p+4\tilde q=3+\tilde q$, exactly as expected.

It is interesting to note that the {\em ordering} of the pasting of the various copies in the recursive construction has no effect on {\bf A}, nor on $M_n$, $N_n$, $\tilde p$, and $\gamma$ .  For example, the rule $\{TQT;QQTT\}$ yields exactly the same results as discussed above for the rule $\{QTT;TQTQ\}$.  The deterministic networks resulting from these two rule sets are different nevertheless --- the difference manifests in other structural properties of the graphs.

\subsection{Statistics of Cycles and Clustering of Mixed Flower Graphs}
\label{ExactLoopiness}

As we have seen, the mixed flower $\{QTT;TQTQ\}$ mimics $S_n([2,3];\tilde p)$, while remaining deterministic.  It therefore affords us an opportunity to explore the statistics of cycles in an exact fashion --- without the replacing of distributions by their averages, used in the analysis of Section~\ref{Loopiness}.

Following a similar notation to that of Section~\ref{Loopiness}, the generating functions for the statistics of paths and loops obey the recursion relations:
\begin{equation}
\label{MixedCycles}
\begin{array}{ll}
\begin{aligned}
\wedge_{n+1} &= \wedge_n^2 + \sqcap_n \\
\triangle_{n+1} & = \wedge_n^2 \sqcap_n + 2\triangle_n+\square_n\\
\sqcap_{n+1} &= \wedge_n\sqcap_n^2  + \wedge_n \\
\square_{n+1} &=  \wedge_n^2 \sqcap_n^2  +  2 \triangle_n + 2 \square_n \\
\end{aligned}
&
\begin{aligned}
;\quad&\wedge_1 = z+z^2,\\
;\quad&\triangle_1 = z^3,\\
;\quad&\sqcap_1 = z+z^3,\\
;\quad&\square_1 = z^4.
\end{aligned}
\end{array}
\end{equation}
(Here we have replaced $\Lt{n}$, $\Ct{n}$, $\Lq{n}$, and $\Cq{n}$ with the more visually obvious notations $\wedge_n$, $\triangle_n$, $\sqcap_n$ and $\square_n$, respectively.)  We stress that the above relations are exact, as the graphs involved are unique and there is no ensemble spread in the statistics of loops of a given length $h$; rather, $\wedge_n(h)$, $\triangle_n(h)$, $\sqcap_n(h)$, and $\square_n(h)$ and their corresponding generating functions are all deterministically determined quantities.

\begin{figure}[ht!]
\centering
\includegraphics[width=4.5in]{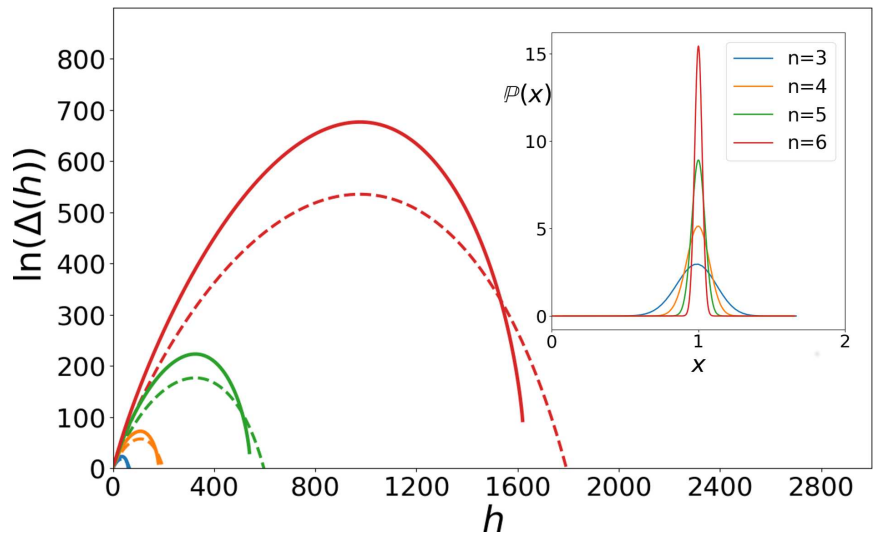}
\caption{(Color online) $\ln\triangle_n(h)$ vs.~the cycle length $h$ for the $\{QTT;TQTQ\}$ flower, as computed from Eqs.~(\ref{MixedCycles}) for generations $n=3,4,5,6$ (solid curves).  For comparison, we also plot the statistics of cycles for the analogous $S_n([2,3];\tilde p)$ flower, as obtained from the approximate Eqs.~(\ref{cyclesSn}) (broken curves, color online).  While disagreement is expected, the identical scaling, of $h_*(n)\sim 3^n$, in both cases, is a nice feature of the approximation.  Inset: The  distribution of $\triangle_n(h/h_*)$ tends to a delta-function as $n\to\infty$.}
\label{MixedLoopy}
\end{figure}

In Fig.~\ref{MixedLoopy} we present results for $\triangle_n(h)$ for generations $n=3,4,5,6$.  Once again, the most likely length for cycles, $h_*$, increases by a factor of 3 from one generation to the next, leading to the same loopiness exponent, $\alpha=\ln 3/\ln\tilde w$, as was found for stochastic $\Sn$ flowers.  For comparison, we also show in the plot the statistics of $C_n(h)$ (broken lines), obtained for $S_n([2,3];\tilde p)$ and the initial conditions of a triangle cycle in generation $n=1$ ($L_1(z)=z+z^2$, $C_1(z)=z^3$).  That the approximate curves for $S_n([2,3];\tilde p)$ and the exact curves for the analogous $\{QTT;TQTQ\}$  do not match comes as no surprise, but it is still reassuring that the approximate approach for stochastic flowers yields the correct scaling and the exact loopiness exponent, nevertheless.

The statistics of cycles provides an example to the importance of the mixed-flowers rule ordering.  Indeed, the rule $\{TQT;QQTT\}$ leads to identical statistics for $M_n$ and $N_n$ as the rule $\{QTT;TQTQ\}$, as previously observed, but to different recursion relations for cycles: for example $\wedge_{n+1}=\wedge_n\sqcap_n+\wedge_n$, instead of the first relation in~(\ref{MixedCycles}), and likewise for the other relations.  This results in the same scaling and loopiness exponent, but  the $\triangle_n(h)$ are not equal in the two cases. 

Clustering in mixed flower graphs can be studied analytically with (exact) recursion relations.  The results are similar to those of deterministic $F_n(1,2)$ and stochastic $\Sn$, with $C_k\sim1/k$ and $\av{C}_n>0$ as $n\to\infty$.  Consider, for example, the clustering coefficient of nodes of degree $k=2$ in a  deterministic graph of generation $n$. These nodes arise only from the iteration of links in generation $n-1$: a link evolving into a triangle yields  one 2-degree node with clustering coefficient $C=1$, while each link evolving into a square yields two 2-degree nodes with $C=0$.  In the thermodynamic limit of $n\to \infty$ the fraction of nodes evolving into triangles (squares) is $\tilde p$ ($1-\tilde p$).  In that case, the average clustering of 2-degree nodes is $C_2\to\tilde p/(2-\tilde p)$.  Since nodes of degree 2 constitute a finite fraction of the nodes in the network, even as $n\to\infty$, the overall clustering coefficient is finite $\av{C}_\infty>0$ (for $\tilde p>0$). 
At the opposite end, consider nodes of degree $k=2^n$ in an $n$-generation network.  For the rule $\{QTT;TQTQ\}$ we get
\begin{equation}
C_{2^n}^\triangle=\frac{2^n+2/3}{2^n(2^n-1)},\qquad C_{2^n}^\square=\frac{2^{n-1}}{2^{n-1}(2^n-1)},
\end{equation}
for nets starting from a triangle or square seed, respectively. In either case, $C_{2^n}^{\triangle,\square}\to 1/(2^n-1)\sim1/k$ as $n\to\infty$.
The details of how $C_k$ converges to its $1/k$ behavior and the actual value for $\av{C}_\infty$ depend on the initial seed and the ordering in the rules of mixing.
Similar remarks can be made for diffusion times between nodes~\cite{Farkas01,Bollt05}, to mention one other obvious example where the ordering of mixing is of consequence.

\section{Conclusion}
\label{Conclusions}

In conclusion, we have explored two different ways to interpolate between the structural properties of $(u,v)$-flower graphs.  {\em Stochastic flowers}  are obtained by selecting $u$ and $v$ in a {\em random} fashion, according to a prescribed set of probabilities.  The closely-related {\em mixed flower graphs}, on the other hand, are obtained by mixing between the rules of growth of different $(u,v)$-flowers in a {\em deterministic} fashion.

Both stochastic and mixed flowers yield themselves to analysis, by exploiting their recursive constructions.  Some structural properties of stochastic flowers, such as their size and order, fail to converge in the thermodynamic limit of infinitely large graphs --- a property we referred to as  ``ensemble spread."  Nevertheless, characteristic exponents of structural properties, such as the degree exponent and loopiness exponent, converge nicely and might be found by replacing the distributions in the recursion relations with their averages, even when the recursion relations are non-linear.

Mixed flowers circumvent the problem of ensemble spread altogether, as there is a unique, deterministic configuration for each mixed flower of any size.  On the other hand, our examples suggest that they mimic stochastic flowers very closely.  It can in fact be shown (future work) that some mixed flowers are members of the $\epsilon$-typical ensemble of stochastic flowers, with respect to the Asymptotic Equipartition Property~\cite{Cover06}. 

For the ease of exposition, we have focused on the simplest examples of stochastic and mixed flowers.  Our examples for each can be generalized in several obvious ways: Stochastic flowers could be obtained by choosing randomly between more than two sets of $(u,v)$ values in each recursive growth.  Mixed flowers, likewise, could be obtained by mixing more than two sets of rules for $(u,v)$-flowers, and a finer gradation in the mixing can be effected by designing rules to produce generation $n+m$ from generation $n$ (with $m>1$, instead of $m=1$).  We have also limited our study to a dearth of structural properties: order, size, degree exponent, statistics of cycles, and clustering.  Many other structural properties, such as assortativity~\cite{Newman03a} and  dynamical properties, such as diffusion between nodes~\cite{Farkas01,Manna03,Bollt05}, can be studied analytically by similar recursive means.

\acknowledgments
EB gratefully acknowledges funding from the Army Research Office (N68164-EG) as well as from DARPA.

\bibliography{Stochastic.bib}

\end{document}